\documentclass[runningheads]{llncs}
\usepackage[T1]{fontenc}
\usepackage{graphicx,verbatim}
\usepackage{amsmath,amssymb}
\usepackage{makecell}
\usepackage{booktabs}
\usepackage{hyperref}
\usepackage{xspace}
\usepackage{bm}
\usepackage{amsmath,amssymb}
\usepackage{multirow}

\newcommand\Tone{ce\text{T}\textsubscript{1}\xspace}

\usepackage[table]{xcolor}
\definecolor{LightGray}{gray}{0.95}

\begin{document}
\title{Deep Biomechanically-Guided Interpolation for Keypoint-Based Brain Shift Registration}
\author{Tiago Assis\inst{1}, Ines P. Machado\inst{2,3}, Benjamin Zwick\inst{4}, Nuno C. Garcia\inst{1}, \and Reuben Dorent\inst{5,6}}
\authorrunning{T. Assis et al.}
\institute{LASIGE, Faculdade de Ciências da Universidade de Lisboa, 1749-016 Lisboa, Portugal\\\email{tassis@lasige.di.fc.ul.pt}\and
CRUK Cambridge Centre, University of Cambridge\and
Department of Oncology, University of Cambridge\and
Intelligent System for Medicine Laboratory (ISML), School of Mechanical Engineering,
The University of Western Australia, Perth 6009, WA, Australia\and
MIND Team, Inria Saclay, Université Paris-Saclay, Palaiseau, France \and
Sorbonne Université, Institut du Cerveau - Paris Brain Institute - ICM, CNRS, Inria, Inserm, AP-HP, Hôpital de la Pitié Salpêtrière, F-75013, Paris, France\\\email{reuben.dorent@inria.fr}
}

\titlerunning{Deep Biomechanical Interpolation for Keypoint-Based Brain Registration}
    
\maketitle              
\begin{abstract}
Accurate compensation of brain shift is critical for maintaining the reliability of neuronavigation during neurosurgery. While keypoint-based registration methods offer robustness to large deformations and topological changes, they typically rely on simple geometric interpolators that ignore tissue biomechanics to create dense displacement fields. In this work, we propose a novel deep learning framework that estimates dense, physically plausible brain deformations from sparse matched keypoints. We first generate a large dataset of synthetic brain deformations using biomechanical simulations. Then, a residual 3D U-Net is trained to refine standard interpolation estimates into biomechanically guided deformations. Experiments on a large set of simulated displacement fields demonstrate that our method significantly outperforms classical interpolators, reducing by half the mean square error while introducing negligible computational overhead at inference time. Code available at: \href{https://github.com/tiago-assis/Deep-Biomechanical-Interpolator}{https://github.com/tiago-assis/Deep-Biomechanical-Interpolator}.


\end{abstract}
\section{Introduction}

Image registration is a fundamental task in image-guided surgery, enabling the spatial alignment of preoperative data with intraoperative anatomy. In neurosurgical procedures, accurate registration is particularly critical, as it supports neuronavigation systems that guide the surgeon based on preoperative Magnetic Resonance Imaging (MRI). However, the reliability of these systems degrades as surgery progresses due to intraoperative brain deformations, often referred to as \textit{brain shift}, caused by gravity, tissue resection, and cerebrospinal fluid loss~\cite{gerard2017}.

To compensate for brain shift, a wide range of registration methods leveraging intraoperative imaging modalities such as intraoperative MRI (iMRI) and ultrasound (iUS) have been proposed, including learning-based~\cite{de2017,balakrishnan2019,mok2020,mok2022} and non-learning-based~\cite{avants2008,ou2011,vercauteren2009} methods. These techniques typically align pre- and intraoperative images by optimizing intensity-based similarity metrics. However, they often struggle in challenging registration scenarios involving (1) large intensity distribution gaps between pre- and intraoperative modalities (e.g., MRI to iUS), (2) large deformations, and (3) topological changes due to tissue resection.

Keypoint-based registration methods have recently gained traction as a competitive alternative~\cite{rister2017,heinrich2022,wang2023,rasheed2024learning}. By relying on sparse correspondences rather than voxel-wise similarity, these methods are more robust to large deformations, partial fields of view, and topological changes. They also offer interpretable outputs, as matched keypoints can be directly visualized and assessed. However, keypoint-based methods typically rely on simple geometric interpolators, such as thin-plate splines or linear models, to propagate sparse displacements into dense displacement fields. These interpolators ignore the biomechanical properties of brain tissue, which can result in physically unrealistic deformations.



\begin{figure}[t!]
\includegraphics[width=\textwidth]{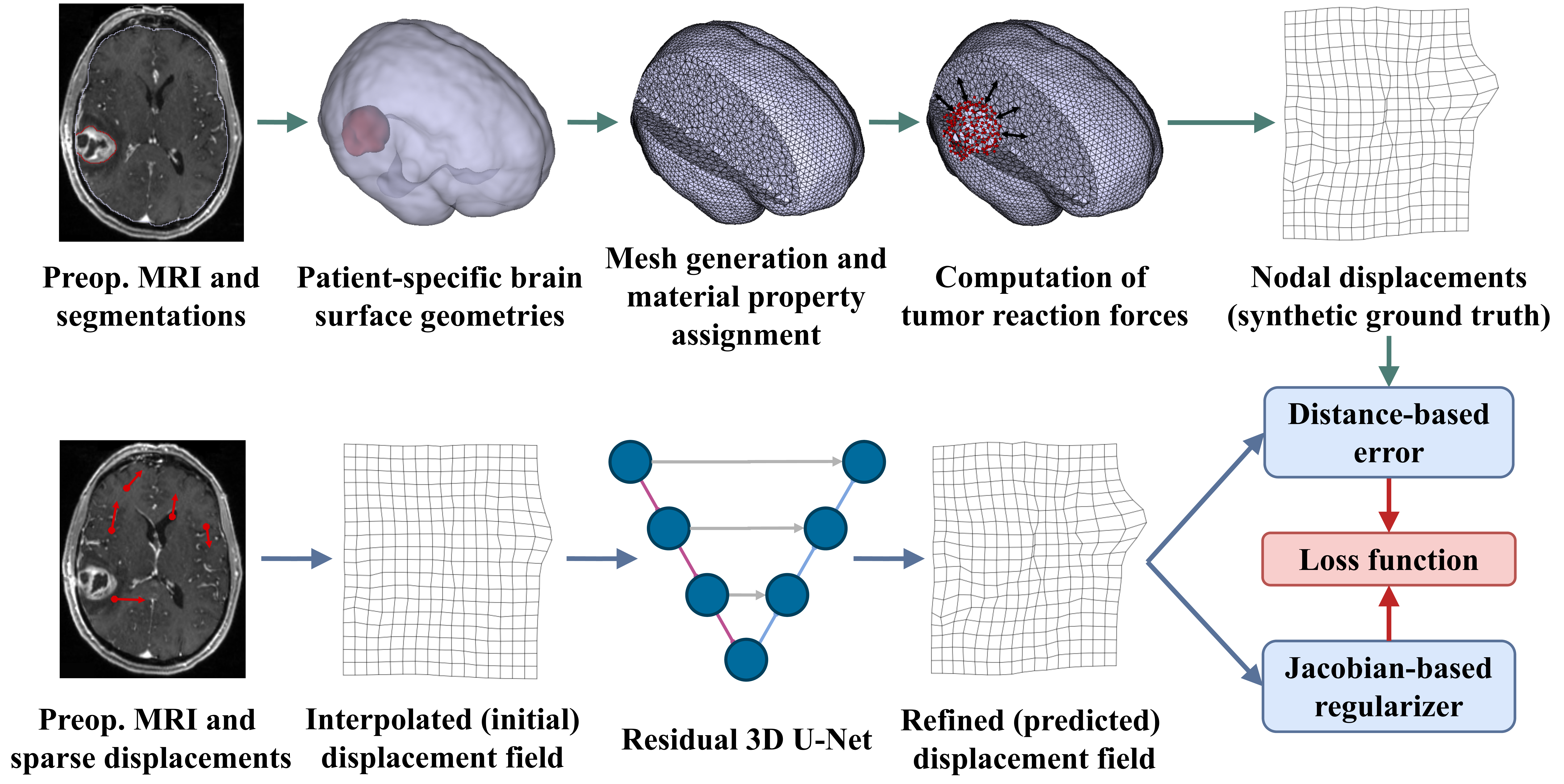}
\caption{Overview of the proposed framework. A biomechanical simulation generates synthetic ground truth displacement fields using preoperative MRI and joint segmentation of tumor and surrounding structures. Sparse intraoperative keypoint displacements are interpolated to form an initial estimate, which is refined by a residual 3D U-Net. The final displacement field is supervised using voxel-wise error and a Jacobian-based regularization loss.} \label{fig:pipeline}
\end{figure}

In this work, we propose a novel deep learning framework for estimating dense and physically plausible brain deformations from sparse matched keypoints between pre- and intra-operative images (Fig.~\ref{fig:pipeline}). First, we construct a large-scale dataset of synthetic brain surgical deformations using biomechanical simulations. Second, we simulate matched keypoints by extracting keypoints using 3D SIFT and pairing them with ground-truth displacements from the synthetic deformations. Third, we develop a deep, biomechanically guided interpolator based on a residual 3D U-Net that refines standard interpolation estimates using the preoperative data. Finally, extensive experiments were conducted on simulated brain deformations, outperforming standard interpolation methods significantly in terms of displacement error, with negligible computational overhead.

\section{Methods}
\subsection{Overview and Problem Setting}
In this work, we assume access to a preoperative MRI $I_{\text{pre}} \in \mathbb{R}^{D\times W \times H}$, where $D$ denotes the depth, $W$ the width, $H$ the height, and a sparse set of $M$ matched keypoints $\{ (\bm{x}_i, \bm{y}_i) \}_{i=1}^M$, where $(\bm{x}_i, \bm{y}_i) \in \mathbb{R}^3\times\mathbb{R}^3$ represent corresponding 3D anatomical locations in the preoperative and intraoperative spaces, respectively. The matched keypoints may be obtained either manually or automatically from any intraoperative imaging modality, such as iMRI or ultrasound. At each keypoint $\bm{x}_i$, the displacement vector $\bm{d}_i = \bm{y}_i - \bm{x}_i$ captures the local brain displacement occurring during surgery. Our objective is to estimate a \textit{dense} and \textit{physically plausible} displacement field $\phi \in \mathbb{R}^{3\times D\times W \times H}$ that estimates the surgical brain deformations. To this end, we propose to train a deep interpolator \( f_\theta \), parameterized by learnable parameters \( \theta \), using biomechanical simulations.

\subsection{Ground-truth Brain Deformation Dataset}
The goal of the deep interpolator $f_\theta$ is to estimate a dense displacement field from a sparse set of $M$ displacement vectors $\{\bm{d}_i\}_{i=1}^{M}$. In the absence of ground-truth dense deformations in clinical data, we propose to train our interpolator using synthetic deformations generated through biomechanical simulations. Specifically, we leverage the biomechanical framework introduced in \cite{yu2022}, which simulates brain deformations induced by tumor resection using the Meshless Total Lagrangian Explicit Dynamics (MTLED) algorithm \cite{joldes2019}. This approach employs a Total Lagrangian formulation with explicit time integration to realistically model intraoperative tissue deformation.

\subsubsection{Brain Tumor Dataset.} We used the UPENN-GBM dataset \cite{bakas2021}, which comprises preoperative multi-parametric brain MRIs and tumor segmentations from $N=162$ patients diagnosed with de novo glioblastoma. For this study, we employed the contrast-enhanced T1 (\Tone) scans as the preoperative MRI $I_{\text{pre}}$, along with their manual segmentations of the tumor core.

\subsubsection{Patient-Specific Geometry.}
The biomechanical framework requires patient-specific brain geometry, including the surfaces of the tumor core and surrounding structures: brain parenchyma, cerebrospinal fluid (CSF), and skull. While the segmentation of brain regions in the presence of tumors can be obtained using dedicated frameworks~\cite{dorent2021learning,himmetoglu2025learning}, these methods do not delineate the CSF and skull. Instead, we use SynthSeg \cite{billot2023}, a tool not specifically designed for pathological data, to segment the parenchyma and CSF on \Tone images. Then, these segmentations are merged with the tumor core segmentation and converted using the ``Model Maker'' module in 3D Slicer \cite{fedorov20123} into triangulated surface models, which are the main input to the biomechanical simulation pipeline.

\subsubsection{Biomechanical Simulations.} Surgical brain deformations are primarily driven by gravity and tissue resection. Following the biomechanical framework introduced in~\cite{yu2022}, we assume that prior to craniotomy, the brain is in an unloaded state in which gravitational forces are balanced by the buoyancy of intracranial fluids. Craniotomy disrupts this equilibrium due to pressure release and CSF drainage, resulting in gravity-induced brain deformation.

The meshless approach uses a cloud of points for spatial discretization and tetrahedral integration cells, simplifying grid construction compared to traditional finite element methods. Two biomechanical brain models are constructed: a pre-resection model that includes the tumor, and a post-resection model in which tumor nodes and their connectivity are removed to simulate the resection cavity. We follow the material parameters for the Ogden model used for Patient 1 in~\cite{yu2022}, which are assigned using fuzzy tissue classification, allowing probabilistic tissue labeling without explicit segmentations.
The parenchyma is modeled as nearly incompressible, the tumor stiffer with a shear modulus three times that of healthy tissue, and CSF as highly compressible to reflect fluid drainage dynamics. 

Then, gravity-induced deformation is simulated by applying the resulting unbalanced forces to the pre-resection model. To model post-resection deformation, internal reaction forces at the tumor-parenchyma interface are computed and applied in the opposite direction to the post-resection model.

Unlike the original framework, where the gravity vector was manually defined, we propose an automated estimation method. Specifically, we derive the base gravity direction from the surface point nearest to the tumor center, based on the assumption that surgeons typically select the shortest access path to the tumor. To account for variability in patient positioning and surgical approach, we generate $K$ plausible gravity vectors by perturbing the base direction by up to $\pm 10^\circ$ along each spatial axis. This leads to the dataset $\mathcal{D}_{\text{total}} = \{ (I^{(j)}_{\text{pre}}, \phi_{\text{gt}}^{(k,j)})_{k=1}^{K} \}_{j=1}^{N}$, containing $K$ distinct displacement fields for each of the $N$ preoperative MRI.

\subsection{Synthetic Matched Keypoints Strategy}
To simulate the acquisition of sparse sets of $M$ displacement vectors $\{ \bm{d}_i \}_{i=1}^M$ for a preoperative MRI $I_{\text{pre}}$, we adopt the following strategy: (1) we extract keypoints from $I_{\text{pre}}$ using the widely used 3D SIFT algorithm \cite{chauvin2020}, and (2) retrieve their associated displacement vectors using the ground-truth synthetic displacement field $\phi_{\text{gt}}$. The 3D SIFT algorithm automatically identifies anatomically meaningful landmarks that are robust to variations in intensity and structure. Since SIFT typically produces hundreds to thousands of keypoints, we randomly sample $M$
keypoints uniformly from the detected set.

\subsection{Deep Physically-Inspired Interpolator}
To design our deep physically-inspired interpolator, we use a denoising approach. Given a sparse set of displacement vectors $\{\bm{d}_i\}_{i=1}^{M}$, we first compute an initial dense displacement field $\phi_{\text{init}}\in \mathbb{R}^{3\times D\times W \times H}$ using a standard interpolation technique such as linear (L) or thin-plate spline (TPS) interpolation. 
This initial estimate $\phi_{\text{init}}$ is then refined by a deep interpolator $f_{\theta}:\left(\mathbb{R}^{D\times W \times H},\mathbb{R}^{3\times D\times W \times H} \right)\mapsto\mathbb{R}^{3\times D\times W \times H}$ conditioned on the preoperative image $I_{\text{pre}}$ to approximate the ground-truth displacement field $\phi_{\text{gt}}$, i.e. $f_{\theta}(I_{\text{pre}},\phi_{\text{init}})\approx \phi_{\text{gt}}$.

\subsubsection{Training Procedure.}
The deep interpolator $f_{\theta}$ is trained under full supervision using the synthetic ground-truth displacement fields. At each training iteration, we randomly sample a training preoperative image $I_{\text{pre}}$ with its 3D SIFT keypoints and a pre-computed ground-truth displacement field $\phi_\text{gt}$. Then, we sample a set of $M$ sparse displacements $\{\bm{d}_i\}_{i=1}^{M}$ and compute on-the-fly an initial dense displacement field $\phi_{\text{init}}$ using a standard interpolation technique (L or TPS). The network $f_{\theta}$ is trained to minimize the mean squared error (MSE) between the predicted $\phi_{\text{pred}}=f_{\theta}(I_{\text{pre}},\phi_{\text{init}})$ and true displacement fields $\phi_\text{gt}$ over the image domain. To encourage smooth displacements in non-resected brain regions, we introduce an additional Jacobian determinant regularization that encourages a local orientation consistency constraint on the estimated displacement field. The total loss function $\mathcal{L}$ then becomes: 
\begin{equation}
    \mathcal{L} \triangleq 
     \| \phi_{\text{pred}} - \phi_{\text{gt}} \|_{2}^2
    + \frac{\lambda_\text{reg}}{|\Omega_{\text{healthy}}|} \sum_{\bm{x} \in \Omega_{\text{healthy}}} \text{ReLU}\left(-\det{J_{I+\phi_{\text{pred}}}}(\bm{x})\right),
\end{equation}
where $\lambda_\text{reg}$ weights the  regularization term, $\Omega_{\text{healthy}}$ denotes the ~\sloppy non-tumorous brain area and $J_{I+\phi_{\text{pred}}}$ is the Jacobian matrix of the deformation field. ReLU ensures that only negative Jacobian determinants are penalized.


\subsubsection{Residual Network Architecture.}
Motivated by denoising diffusion models~\cite{ho2020denoising}, which have shown that learning to predict the noise in a noisy signal leads to better performance than predicting the clean signal directly, our network predicts a residual displacement $\epsilon_\theta$, such that $f_\theta(I_{\text{pre}}, \phi_{\text{init}}) = \phi_{\text{init}} + \epsilon_\theta(I_{\text{pre}}, \phi_{\text{init}})$. The residual network $f_\theta$ is a 3D U-Net architecture variant of \cite{lee2017}. At each resolution level, residual blocks are employed with spatial and channel-wise squeeze-and-excitation (SE) \cite{roy2018} modules. 
Downsampling in the encoder path is performed using max-pooling, while upsampling in the decoder path is achieved via transposed convolutions. Same-size feature maps from the encoder are merged with decoder features through element-wise summation rather than concatenation. The network comprises $4$ resolution levels, starting with $32$ feature channels and doubling at each downsampling stage up to a maximum of $256$ channels, while the spatial resolution is halved. Each convolution within the residual blocks is followed by instance normalization and a LeakyReLU activation function with a negative slope of $10^{-2}$, except for the final convolution in each block, where activation is applied after the residual summation. The SE modules are then applied at the end of each block, resulting in a network with 7.3M parameters.

\begin{figure}[t!]
\includegraphics[width=\textwidth]{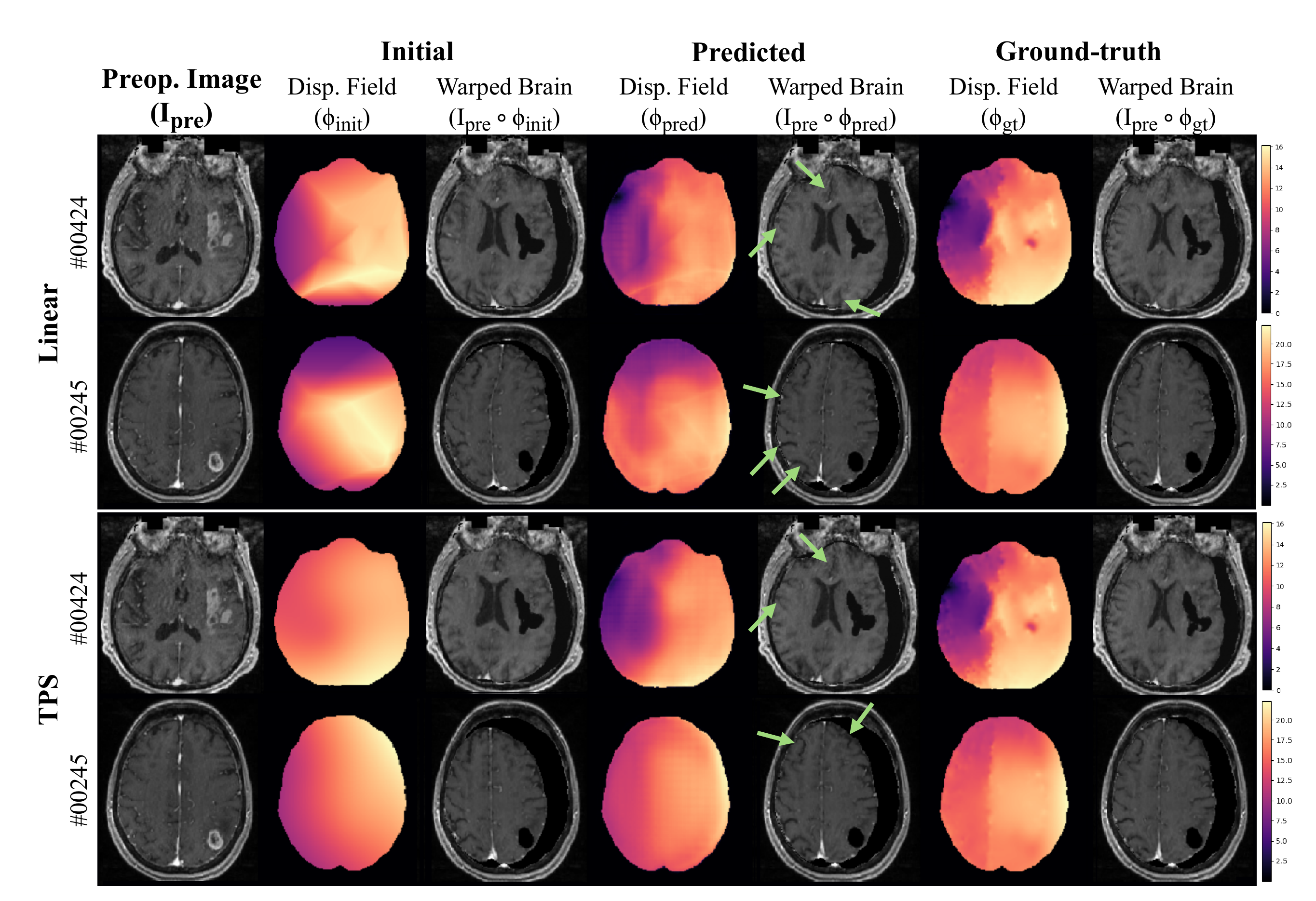}
\caption{Qualitative comparison of displacement fields and resulting warped brain anatomies. The displacement fields are colored by vector magnitudes, and green arrows highlight the most noticeable improvements over the linear and TPS baselines. 
} \label{results_brains}
\end{figure}

\section{Experiments}

\subsubsection{Data.}
We evaluated our method using the UPENN-GBM dataset, where 204 synthetic ground-truth displacement fields were generated via biomechanical simulations for 162 unique patients ($K=1$-3 simulations per case). The dataset was split into 121 training, 16 validation, and 25 test cases, following a 75:10:15 ratio. For each case, a ground-truth displacement field was randomly selected from the set of simulations with different gravity-induced brain shifts, and a random set of keypoints was sampled to initialize the displacement field.

\subsubsection{Implementation.}
All inputs were cropped to a fixed size of $160\times192\times144$.
Preoperative \Tone images were normalized by subtracting the mean and dividing by the standard deviation. Data augmentation was applied only to images during training and included Gaussian noise and blur, intensity adjustments (brightness, contrast, gamma), and simulated low resolution. These augmentations followed the strategies used in nnU-Net \cite{isensee2021} to improve generalization capability. Our approach was trained using the Adam optimizer with a learning rate of $5\times10^{-4}$, a batch size of $1$, and for 100 epochs. To ensure anatomical relevance for all methods, interpolated displacements in the background or skull are set to zero. A value of $\lambda_\text{reg} = 50$ was chosen by performing a grid search on the validation set, which provided the best balance between performance and regularization.

\begin{table}[t!]
\centering
\begingroup
\caption{Quantitative evaluation of different approaches using linear (L) and thin-plate spline (TPS) interpolation using $M=20$ keypoints. R denotes the residual architecture and J the Jacobian regularization term. Mean and standard deviation are reported. Statistical significance was determined using a Bonferroni-corrected paired Wilcoxon signed-rank test, * indicating statistically significant improvements (p < 0.01) for Ours. }
\label{tab:baselines_ablations}
\fontsize{8}{10}\selectfont
\begin{tabular}{lcccccc}
\toprule
\multirow{2}{*}{\textbf{Method}} & \multicolumn{2}{c}{\textbf{MSE} (mm$^2$)$\downarrow$} & \multicolumn{1}{c}{\textbf{Max Error}$\downarrow$} & \multicolumn{1}{c}{\textbf{HD95}$\downarrow$} & \multicolumn{1}{c}{$\%|\textbf{J}_\phi|\textbf{<0}$$\downarrow$} & \multicolumn{1}{c}{\textbf{Time}$\downarrow$} \\
\cline{2-3}
& Brain & Edema & (mm) & (mm) & (\%) & (s) \\
\midrule
\rowcolor{LightGray}
L (baseline)        & 10.7* (5.4) & 7.6 (5.6) & 28.3 (8.0) & 3.7* (1.1) & 0.74* (0.11) & 1.81 (0.02) \\
$\text{Ours (L)}_{\text{w/o R+J}}$            & 4.2 (2.2) & 10.9* (6.6) & 27.8 (8.5) & 2.8 (0.7) & 0.97* (0.20) & - \\
\rowcolor{LightGray}
$\text{Ours (L)}_{\text{w/o J}}$        & 3.7 (1.7) & 9.4 (7.1) & 27.3 (8.4) & 2.8 (0.5) & 1.16* (0.25) & - \\
Ours (L)    & 3.7 (1.6) & 6.4 (3.0) & 26.2 (8.2) & 2.7 (0.6) & 0.64 (0.21) & 1.81 (0.02) \\
\hline
\rowcolor{LightGray}
TPS (baseline)        & 6.5* (2.6) & 6.4 (5.5) & 25.4* (6.9) & 2.8 (0.7) & 0.64* (0.21) & 0.58 (0.01) \\
$\text{Ours (TPS)}_{\text{w/o R+J}}$            & 4.6* (2.1) & 10.6* (7.5) & 26.4 (8.7) & 2.9 (1.0) & 1.47* (0.36) & - \\
\rowcolor{LightGray}
$\text{Ours (TPS)}_{\text{w/o J}}$        & 3.5 (1.6) & 7.2 (5.2) & 23.4 (5.2) & 3.3 (0.7) & 0.99* (0.14) & - \\
Ours (TPS)     & 3.4 (1.6) & 5.9 (3.3) & 22.7 (4.9) & 3.1 (0.5) & 0.59 (0.22) & 0.59 (0.01) \\
\bottomrule
\end{tabular}
\endgroup
\end{table}

\subsubsection{Metrics.}
To assess the performance of our method, we evaluated the predicted displacement fields using several complementary metrics. The mean squared error (\textbf{MSE}) in mm$^2$ was computed between the predicted and ground-truth displacement fields, both within the whole brain and specifically within the edematous tumor region, to quantify the overall accuracy and the accuracy near the resected area, respectively. We additionally reported the maximum Euclidean error (\textbf{Max Error}) in mm, capturing the worst-case deviation in the predicted displacements. To assess the geometric alignment of brain structures, we computed the 95th percentile Hausdorff distance (\textbf{HD95}) between the brain segmentations warped by the predicted and ground-truth displacement fields. To evaluate the anatomical plausibility of the deformations, the percentage of voxels with non-positive Jacobian determinant values ($\bm{\%|J_\phi|<0}$) was computed. Finally, we reported the inference time (\textbf{Time}) to evaluate the computational efficiency of our method in comparison to the baselines.

\subsubsection{Baseline Interpolation Methods and Ablation Study.} Two widely-used interpolation methods were used in our experiments for comparison and initialization: (1) linear interpolation (L) via a 3D Delaunay triangulation approach using a publicly available differentiable implementation~\cite{joutard2022} and (2) a thin-plate spline (TPS) approach using a public implementation \cite{wang2023}. A regularization weight $\lambda_\text{tps}$ of 0.1 was used for TPS, which achieved the best empirical results on the validation set.
Both interpolation methods serve as non-learning-based baselines that do not incorporate anatomical context or physical priors.

We also performed an ablation study to evaluate the impact of three key components of our method: (1) the choice of interpolator for displacement field initialization, (2) a residual network architecture (R), and (3) the Jacobian regularization term (J). All configurations used a fixed set of 20 keypoints per case.



\subsubsection{Results.} Results are shown in Table~\ref{tab:baselines_ablations} and Fig.~\ref{results_brains}.
Our method consistently outperformed baseline interpolators, reducing whole-brain MSE by up to $47\%$ ($3.07$ mm$^2$) with TPS and $65\%$ ($6.96$ mm$^2$) with linear interpolation. Improvements in the edema were more modest but with reduced variability, suggesting a more stable approach. The residual architecture improved accuracy across the board, especially in the edema region, by enabling the model to learn finer corrections. Notably, omitting residual learning with TPS initialization led to increased voxel folding, which can be attributed to TPS's lack of flexibility to adapt to fine-grained deformations, leading the network to overcompensate. Adding a Jacobian regularizer improved deformation smoothness, reducing non-invertible mappings by $40$ to $45\%$ ($+0.40$ to $+0.52$pp), without compromising accuracy. Inference time increased negligibly compared to baseline methods ($+10$ms).




\begin{table}[t!]
\centering 
\begingroup
\caption{Impact of the number of $M$ matched keypoints in terms of MSE (mm$^2$).}
\label{tab:num_keypoints}
\fontsize{8}{10}\selectfont
\begin{tabular}{lccccc}
\toprule
\multirow{1}{*}{\textbf{Method}} & \multicolumn{1}{c}{$M=5$} & \multicolumn{1}{c}{$M=10$} & \multicolumn{1}{c}{$M=15$} & \multicolumn{1}{c}{$M=20$} & \multicolumn{1}{c}{$M=50$} \\
\midrule
\rowcolor{LightGray}
L (baseline)  &  17.2 (9.1) & 13.3 (6.5) & 11.4 (5.9) & 10.7 (5.4) & 6.4 (3.1) \\
Ours (L)    &  \textbf{7.8 (4.3)} & \textbf{4.8 (2.5)} & \textbf{4.7 (2.7)} & \textbf{4.0 (2.0)} & \textbf{2.3 (1.1)} \\
\hline
\rowcolor{LightGray}
TPS (baseline) &   18.0 (12.4) & 10.2 (4.8) & 7.2 (2.9) & 6.5 (2.6) & 3.8 (1.6) \\
Ours (TPS)   &   \textbf{11.0 (8.4)} & \textbf{5.7 (3.3)} & \textbf{4.4 (2.5)} & \textbf{3.4 (1.6)} & \textbf{2.1 (1.1)} \\
\bottomrule
\end{tabular}
\endgroup
\end{table}

\subsubsection{Impact of Number of Keypoints.}
Finally, we analyzed the impact of the number of input keypoints $M$, varying it from 5 to 50 (Table~\ref{tab:num_keypoints}). As expected, increasing the number of keypoints led to lower MSE, as the interpolation benefits from more accurate and localized displacement observations. At low keypoint counts (e.g., $5$), TPS interpolation performed poorly, likely due to instability with limited control points. In contrast, linear interpolation demonstrated greater robustness in such settings. However, with a higher number of keypoints (e.g., $50$), TPS produced smoother and more accurate results, outperforming linear interpolation. This illustrates a trade-off between robustness and smoothness that depends on the spatial density of keypoints. Notably, in all cases, our deep interpolator significantly improved upon the initial interpolation, reducing error regardless of the number of points $M$ and the interpolation method.

\section{Conclusion} 
We introduced a deep learning framework for estimating dense and physically plausible brain deformations from sparse keypoint correspondences between pre- and intra-operative images. To enable supervised training, we built a large dataset of synthetic brain deformations using biomechanical simulations across a large number of cases. Sparse keypoints were simulated using 3D SIFT and paired with ground-truth displacements to mimic intraoperative correspondences. A residual 3D U-Net was trained to refine standard interpolation fields into biomechanically-guided deformations, guided by the preoperative image and regularized by a Jacobian-based constraint. Our experiments demonstrate that the proposed method consistently outperforms classical interpolators in accuracy, without incurring significant computational cost at inference. Future work will explore the generalization capabilities of our framework on additional datasets, such as the ReMIND dataset~\cite{juvekar2024remind}. We also aim to extend the method's robustness to handle imperfect correspondences and varying numbers of matched keypoints. Finally, we plan to apply our approach to image registration for surgical guidance.

\begin{credits}
\subsubsection{\ackname} Tiago Assis received a grant from Fundação para a Ciência e a Tecnologia (FCT), I.P./MCTES, through national funds (PIDDAC) under the Strategic Funding 2020-23 programme (Grant UIDB/00408/2020). Reuben Dorent received a Marie Skłodowska-Curie grant, No. 101154248 (project: SafeREG). This activity has been supported by the Western Australian Future Health Research and Innovation Fund (Grant ID WANMA/Ideas2023-24/13). 

\subsubsection{\discintname}
The authors have no competing interests to declare that are
relevant to the content of this article.
\end{credits}

%
%
%

\bibliographystyle{splncs04}
\bibliography{refs}

\end{document}